\documentclass[11pt,dvips,ismdproc]{article}
\usepackage{cite,mcite}
\usepackage{graphicx}
\usepackage{subfigure}
\usepackage{mathptmx}
\usepackage{helvet}
\usepackage{multirow}
\newlength{\capindent}
\newlength{\capwidth}
\newlength{\figwidth}
\newcommand{\icaption}[2][!*!,!]{\hspace*{\capindent}%
  \begin{minipage}{\capwidth}
    \ifthenelse{\equal{#1}{!*!,!}}%
      {\caption{#2}}%
      {\caption[#1]{#2}}
      \vspace*{3mm}
  \end{minipage}}
\begin{document}
%
\title{Tuning of the helix string model \\
 on the DELPHI $Z^0\rightarrow q\bar{q}$ data}  

\author{\v{S}.Todorova}

\maketitle
%
%
\begin{abstract}
     
    An alternative model of the fragmentation of the Lund string, investigating
 the hypothesis of helix-like ordered gluon field, is compared with hadronic $Z^0$ data.
 A significant improvement in the description of various measured quantities is achieved.
 In particular, the existence of correlations between longitudinal and transverse components
 of hadron momenta (as suggested by the helix string model) seems to be supported by the data.
  
\end{abstract}
%
%
%
\section{Introduction}

   The possibility that a helix-like ordered gluon field emerges at the end of a parton cascade 
 was first discussed in \cite{Andersson:1998sw}, on the basis of a study of optimal
 packing of soft gluons in the phase space. In their pioneering paper, the authors suggested
 to study a helix parametrization related to the rapidity difference along string.
 The search of such a helix string signature was performed but no significant signal
 was found \cite{delphi_scr}.

   In this paper, we will study an alternative parametrization of the helix string, which
 may be at the origin of some persistant discrepancies between data and modelling.  
 
 The paper is organized as follows :  in section 2, a very brief description of the model is given
 ( for details see \cite{ismd_helix} ). Section 3 deals with the implementation of the model. Section
 4 describes the tuning setup and strategy. Results of tuning are summarized in Section 5. 

\clearpage

 \section{(Modified) helix string model} 

  Assuming the string field can be modelled by a stream of soft gluons (Fig.~\ref{fig:helix}),
 direct hadrons acquire their transverse momenta by \emph{integration}
  over soft gluon momenta  along the field:
\begin{figure}[h]
\begin{minipage}{0.5\textwidth}
  \begin{equation}
     | \vec{p_T}  | = 2 r | sin \frac{\Delta\Phi}{2}  |
 \end{equation}
   where $\vec{p_T}$ is the transverse momentum of the hadron, $r$ stands for the radius of the helix, and $\Delta\Phi$ is the difference of the helix phase between the string break-ups which define the hadron. 

 In the modified helix model, the helix winding is proportional to the energy density
 of the string: 
\end{minipage}    
\hspace{0.1\textwidth}    
\begin{minipage}[h]{0.4\textwidth}

\begin{flushright}
\includegraphics[width=1.\textwidth]{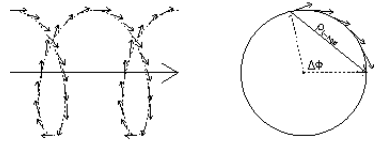}
\vspace{-0.5cm} 
\caption{ Helix-like ordering of soft gluons at the end of parton cascade (left) and direct hadron transverse momentum (right).}
\label{fig:helix}
\end{flushright}

\end{minipage}
\end{figure}
 
    \begin{equation}
\label{eq:myphase}
   \Delta\Phi = S \ \kappa  \ \Delta l = S \ E_{had} 
\end{equation} 
   where $\kappa$ stands for the string tension [GeV/fm], $\Delta l$ is the length of string piece, and $S$ [rad/GeV] is a parameter. $E_{had}$ corresponds to the energy of the direct hadron in the string c.m.s. Parametrization (\ref{eq:myphase}) implies strong correlation between energy and transverse momentum of the direct hadron (in the rest frame of the string).

\section{Implementation of  the model}

   The modified helix model is implemented in Pythia 6.421 \cite{Sjostrand:2006za} via private version of  the fragmentation routine
 PYSTRF \cite{pystrf}.  The following Pythia parameters and switches are used for steering:

\begin{itemize}
\item{MSTU(199)=} 0/2 (standard fragmentation/modified helix fragmentation)
\item{PARJ(102)=} * ( helix radius r [GeV/c], replaces PARJ(21) )
\item{PARJ(103)=} * ( helix radius variance [GeV/c] )
\item{PARJ(104)=} * ( parameter S [rad/GeV] )
\item{PARJ(109)=} 0.001 (azimuthal angle tolerance in the iterative search of the string break-up solution
 conform to Lund fragmentation rules \emph{and} helix string parametrization)  
\end{itemize}


\section{Tuning setup and strategy}

    The tuning is performed using the Rivet and Professor packages \cite{Buckley:2009bj}, using
 alternatively Pythia $p_T$ ordered shower \cite{Sjostrand:2006za}, and Ariadne parton shower \cite{ariadne},
 on top of hadronic $Z^0$ decay generated by Pythia 6.421.

 The set of 6 simultaneously
 tuned parameters consists of
\begin{itemize}
\item{ helix radius}  - PARJ(102)
\item{ helix winding}  - PARJ(104)
\item{ Lund parameter \emph{a}}   - PARJ(41)
\item{ Lund parameter \emph{b}}  - PARJ(42)
\item{ effective coupling constant $\Lambda_{QCD}$}  - PARJ(81)
\item{ parton shower cut-off}  - PARJ(82)
\end{itemize}
  (the later 2 parameters are replaced by PARA(1) and PARA(3) in the Ariadne tune).

    The setup of other Pythia parameters is based on the Professor tune of $p_T$ ordered parton shower \cite{prof_tune}, or on the
  DELPHI tuned setup \cite{tuning}, with small modifications described in the Appendix.

    For simplicity, the helix radius variance is fixed (PARJ(103)=0.1 GeV/c). 

    The tuning is performed on the set of inclusive charged particle spectra and event shape distributions measured by the DELPHI Collaboration \cite{tuning}. 

    The input samples, generated with Rivet and private Pythia library, contain 500k events each. The set of fitted
  optimal parameters is used to generate a validation sample which is then compared to the reference data (i.e., the reported $\chi^2$ difference refers to the validated result of the tune
 rather than interpolated estimate).

\section{Results and comparison with data}
  
\begin{figure}[bth]
\includegraphics[width=\textwidth]{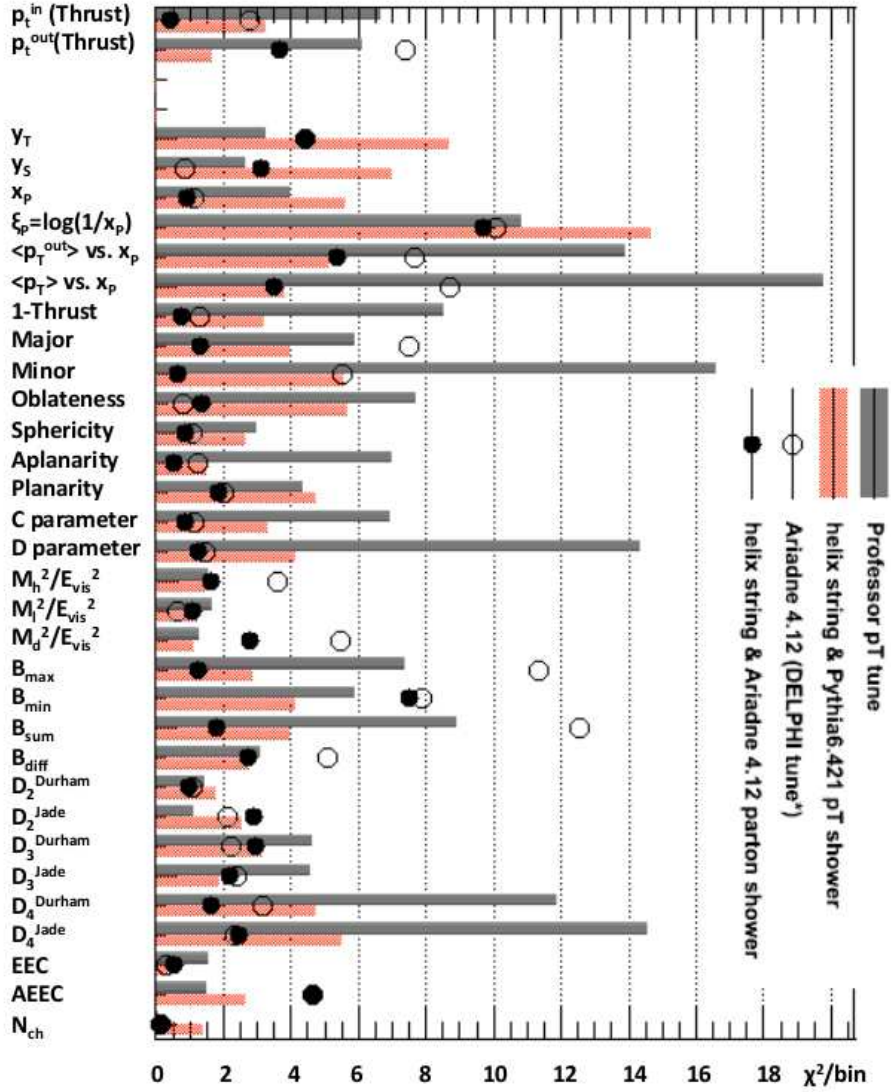}
\caption{Comparison of tuned model predictions with DELPHI data, in $\chi^2$ per bin.
 Bar charts stand for modelling with Pythia $p_T$ ordered shower, points mark modelling
 with Ariadne parton shower. On average, helix string model (dashed chart and closed points)
 provides a better description of the DELPHI data.}
\label{fig:chi2perbin}
\end{figure}

   Data distributions included in the fit are listed in Fig.\ref{fig:chi2perbin} (for
 definitions see \cite{tuning}). The comparison of data with tuned model predictions
 is quantified by $\chi^2$ per bin measure, separately for each observable:
\begin{equation}
     \chi^2/bin = \frac{1}{N_{bin}} \Sigma_{bin} \frac{ (X_{data}-X_{MC})^2} { \sigma(X)_{data}^2}
\end{equation}

\clearpage

\begin{figure}[ht]
\includegraphics[width=0.5\textwidth]{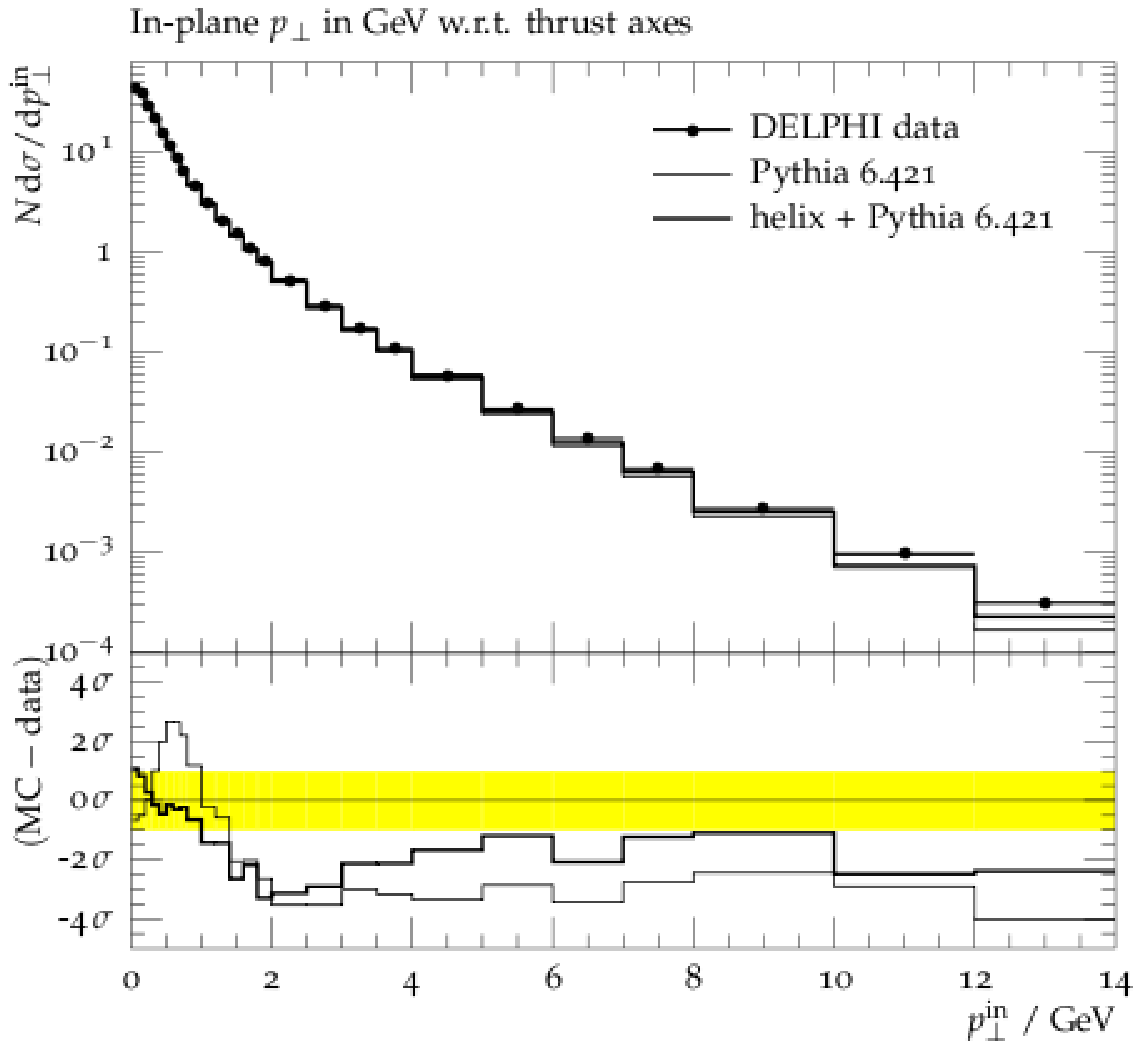}
\includegraphics[width=0.5\textwidth]{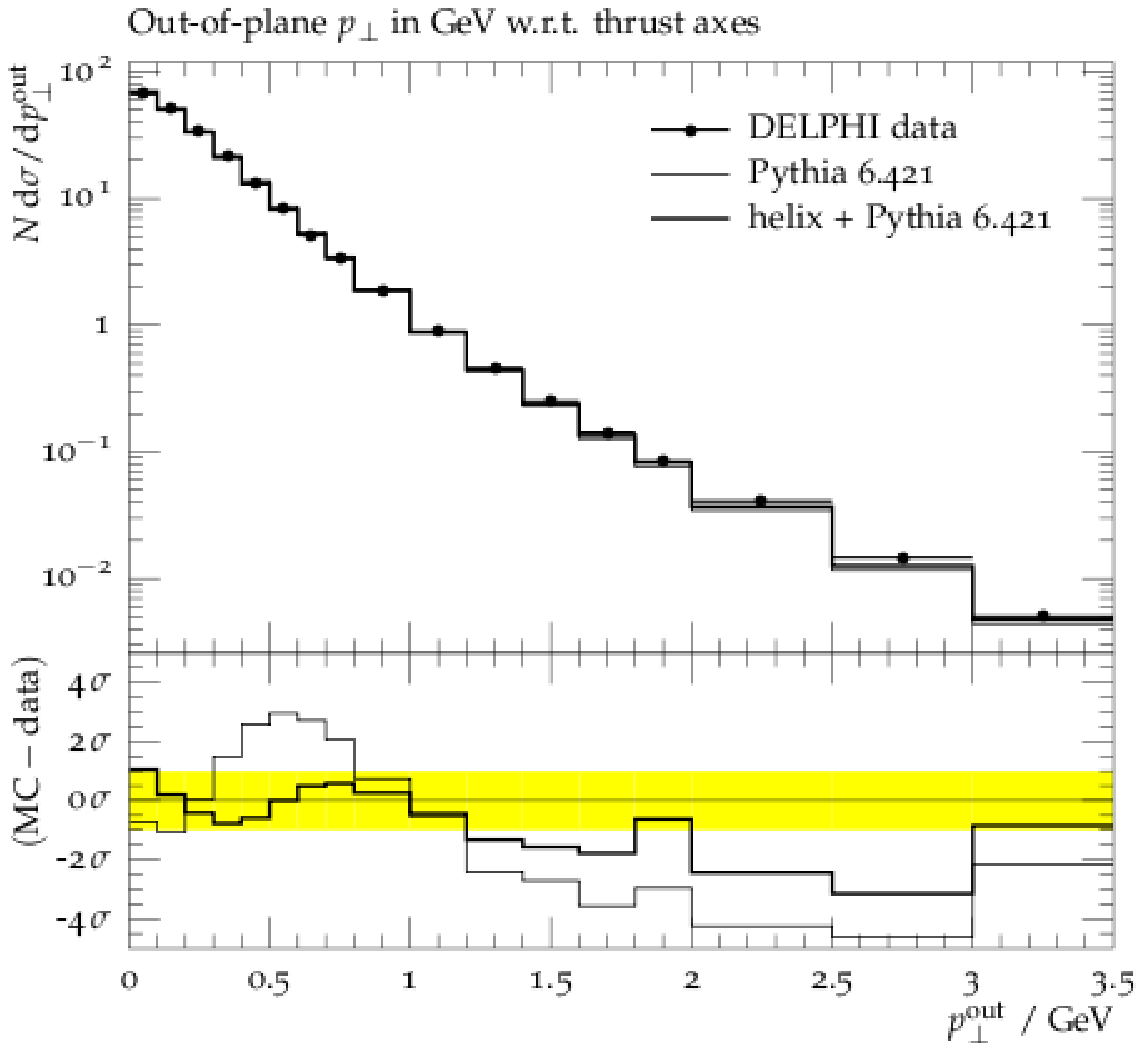}
\includegraphics[width=0.5\textwidth]{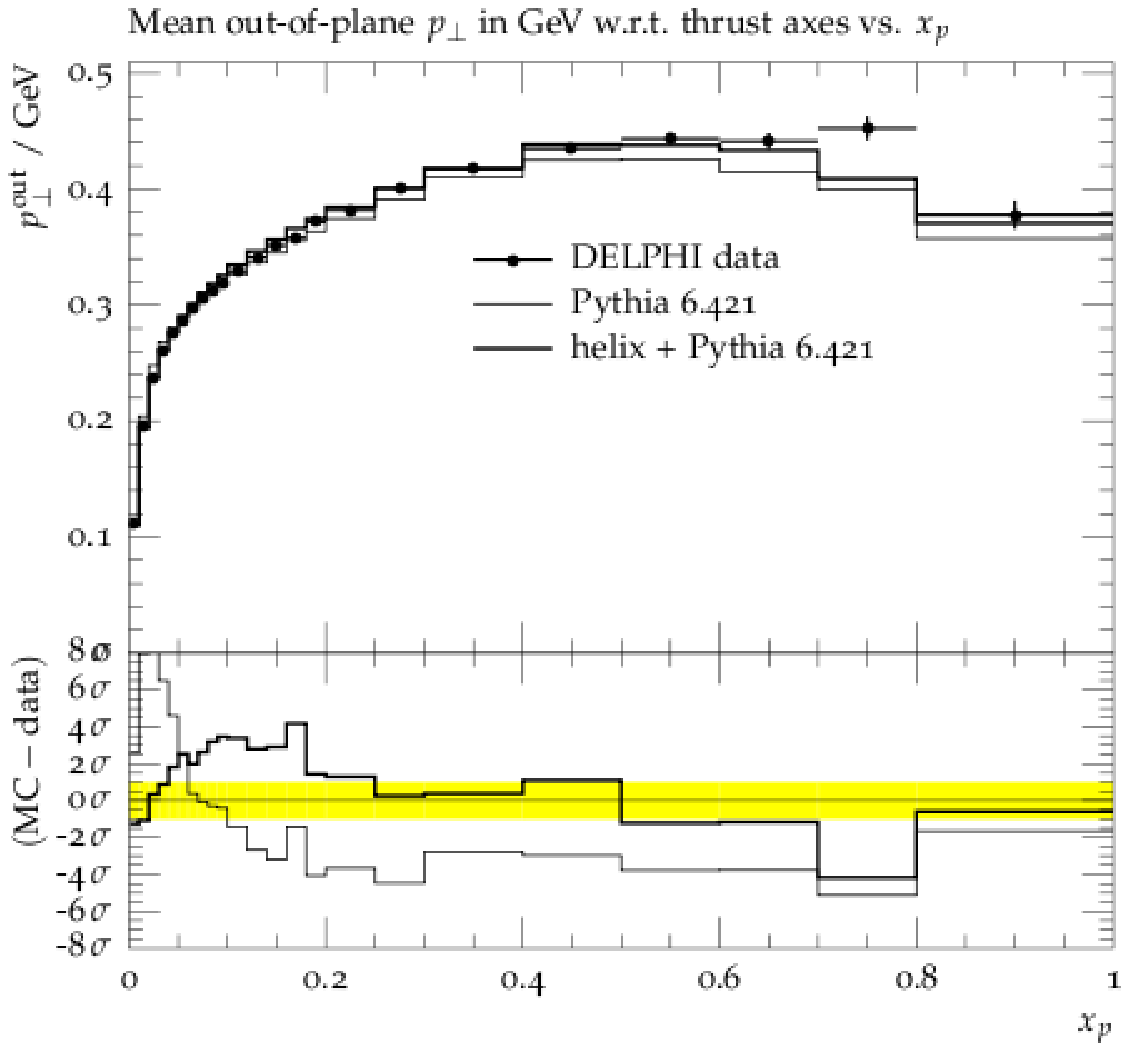}
\includegraphics[width=0.5\textwidth]{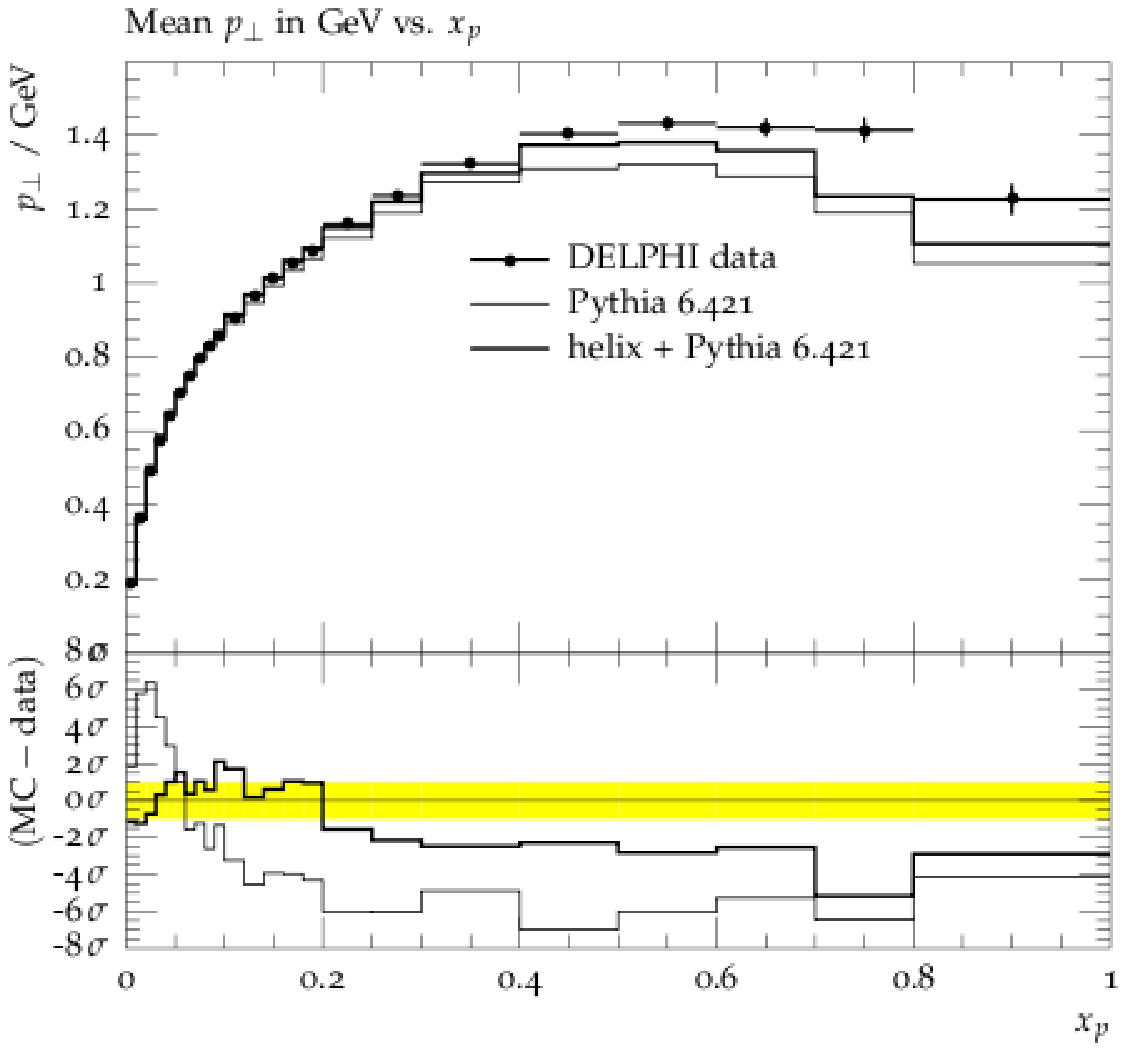}
\includegraphics[width=0.5\textwidth]{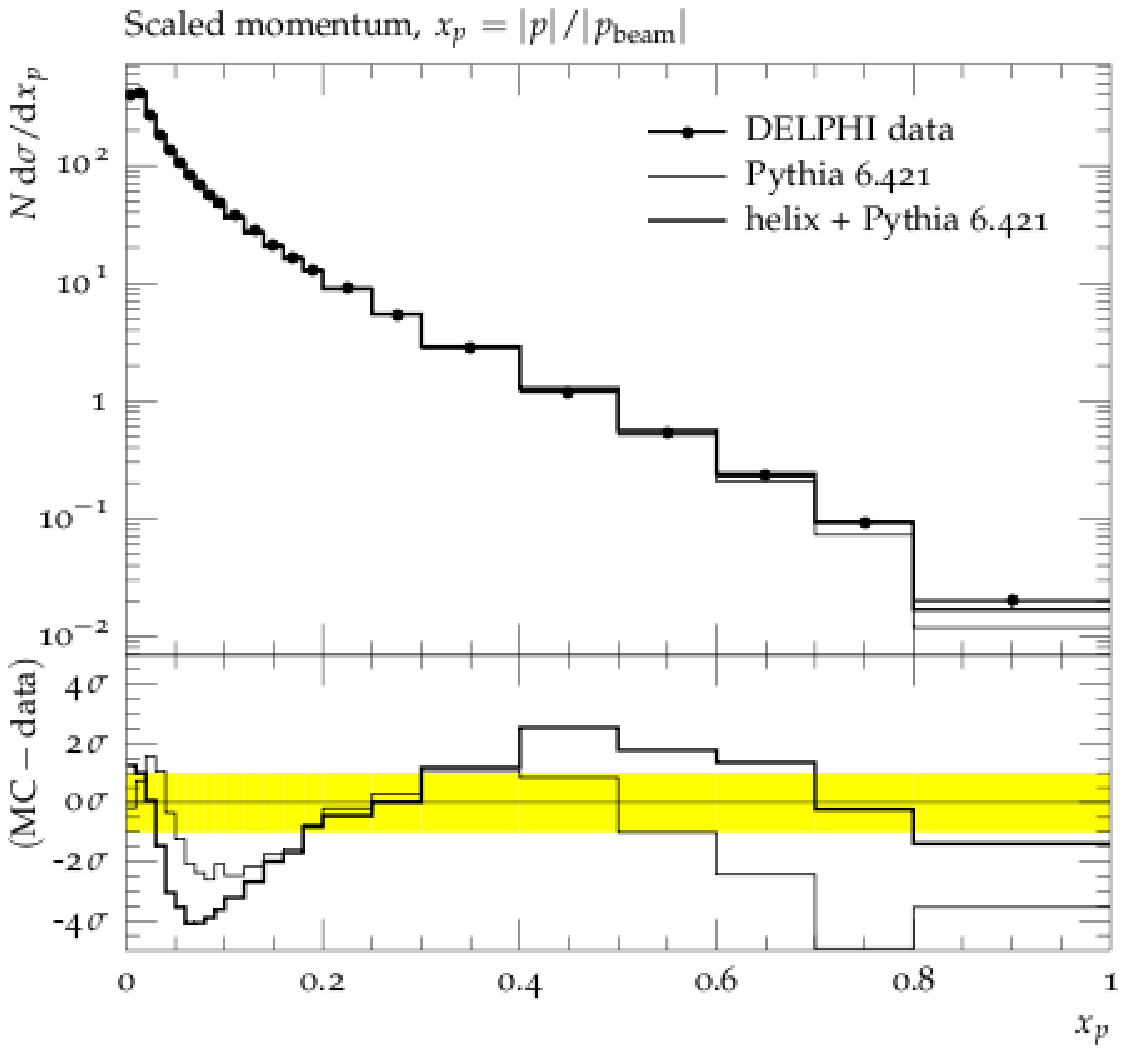}
\includegraphics[width=0.5\textwidth]{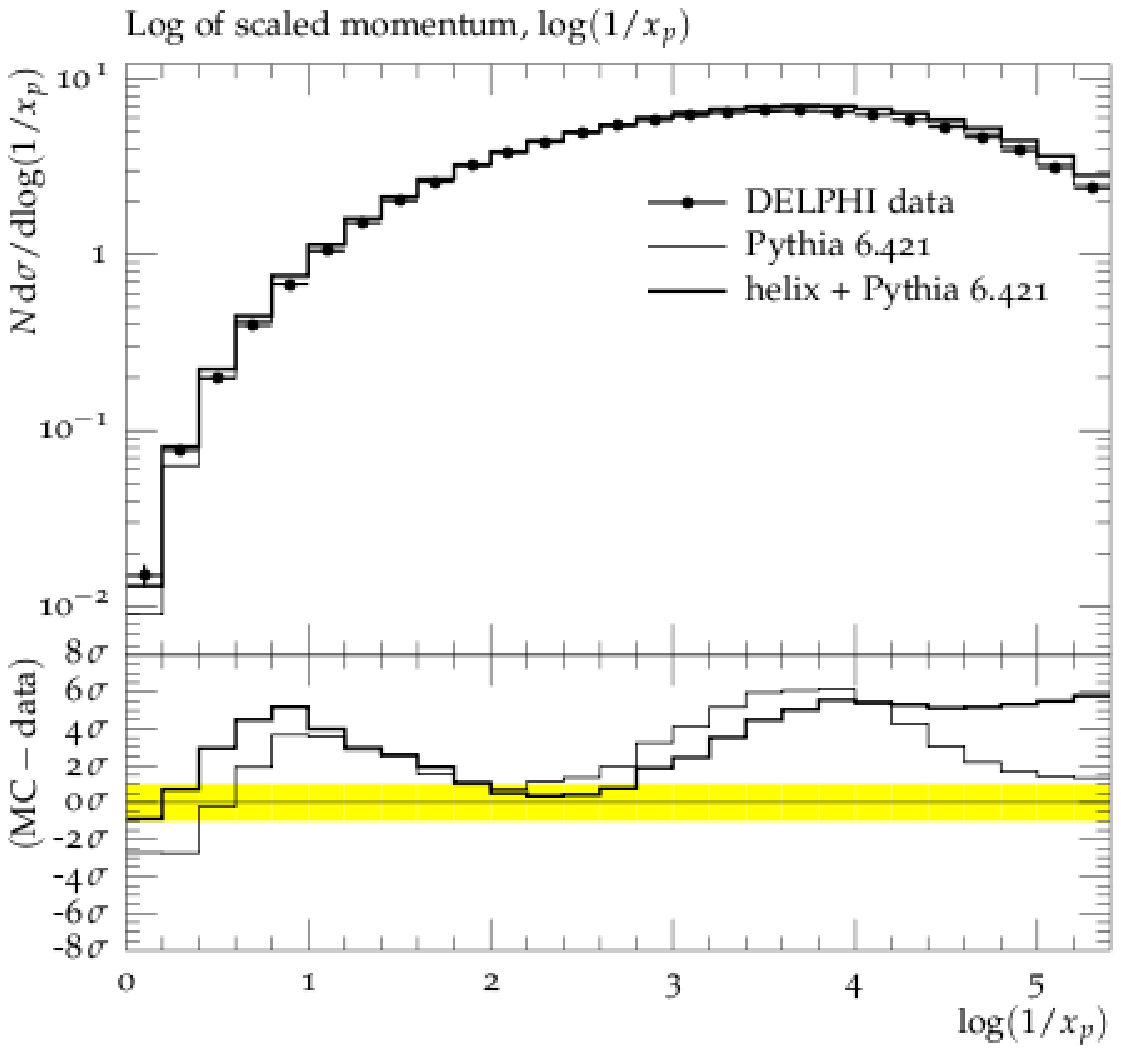}
\caption{Comparison of inclusive charged particle distribution measured by DELPHI
 and MC simulation based on Pythia pT-ordered shower, using standard
 fragmentation ('Pythia 6.421') or helix string model ('helix+Pythia 6.421').}
\label{fig:incl_pt} 
\end{figure}

\begin{figure}[ht]
\includegraphics[width=0.5\textwidth]{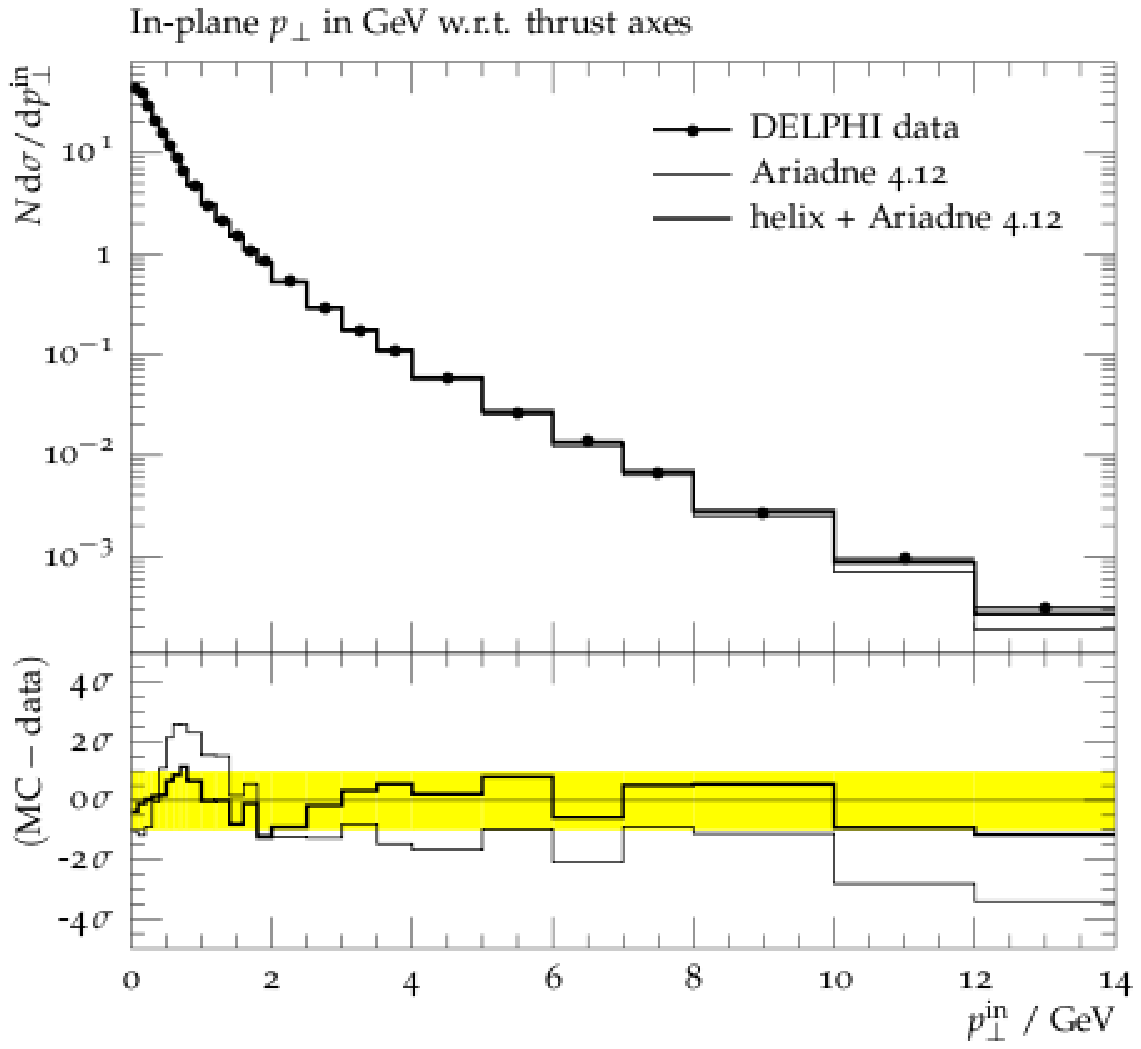}
\includegraphics[width=0.5\textwidth]{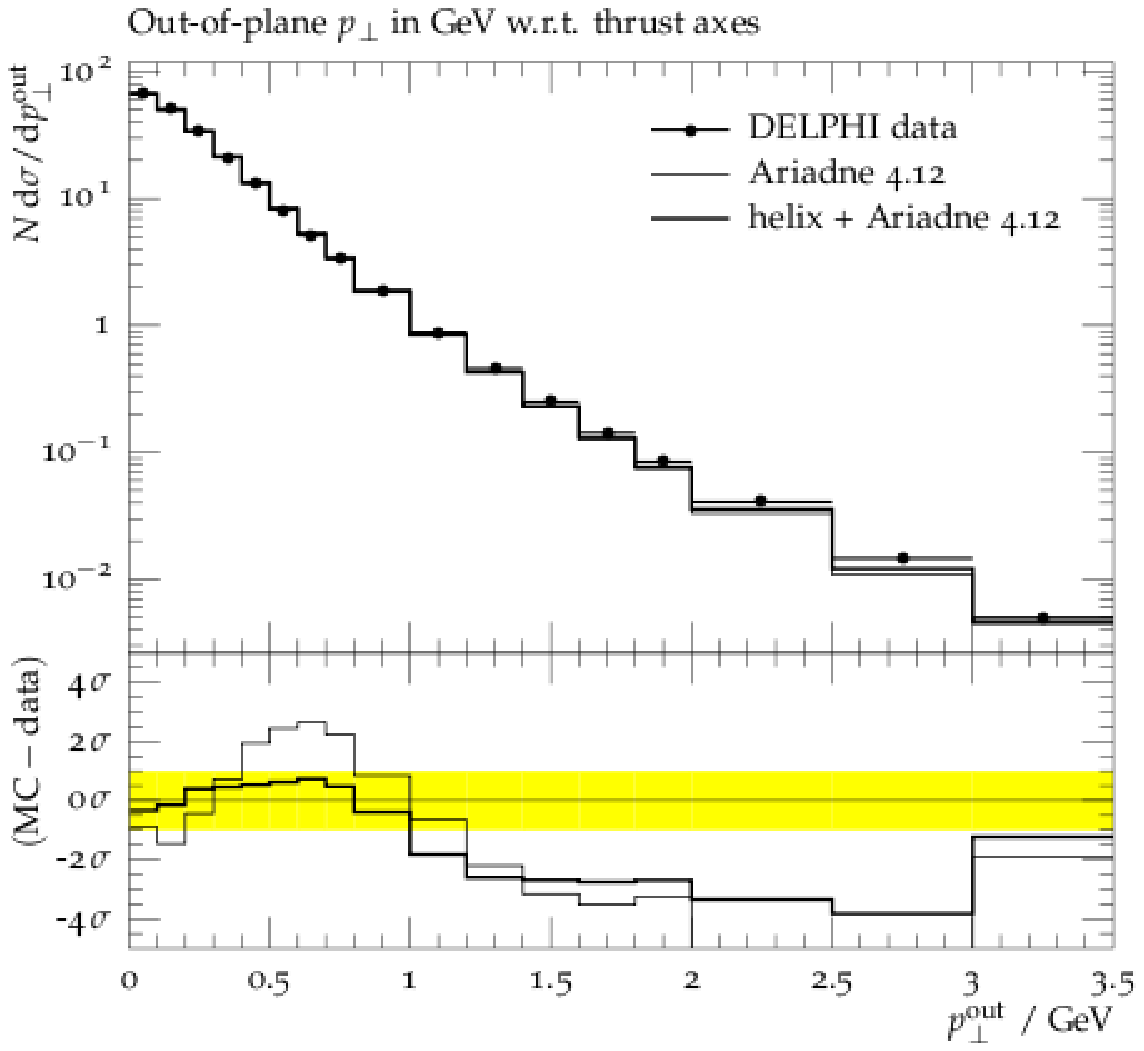}
\includegraphics[width=0.5\textwidth]{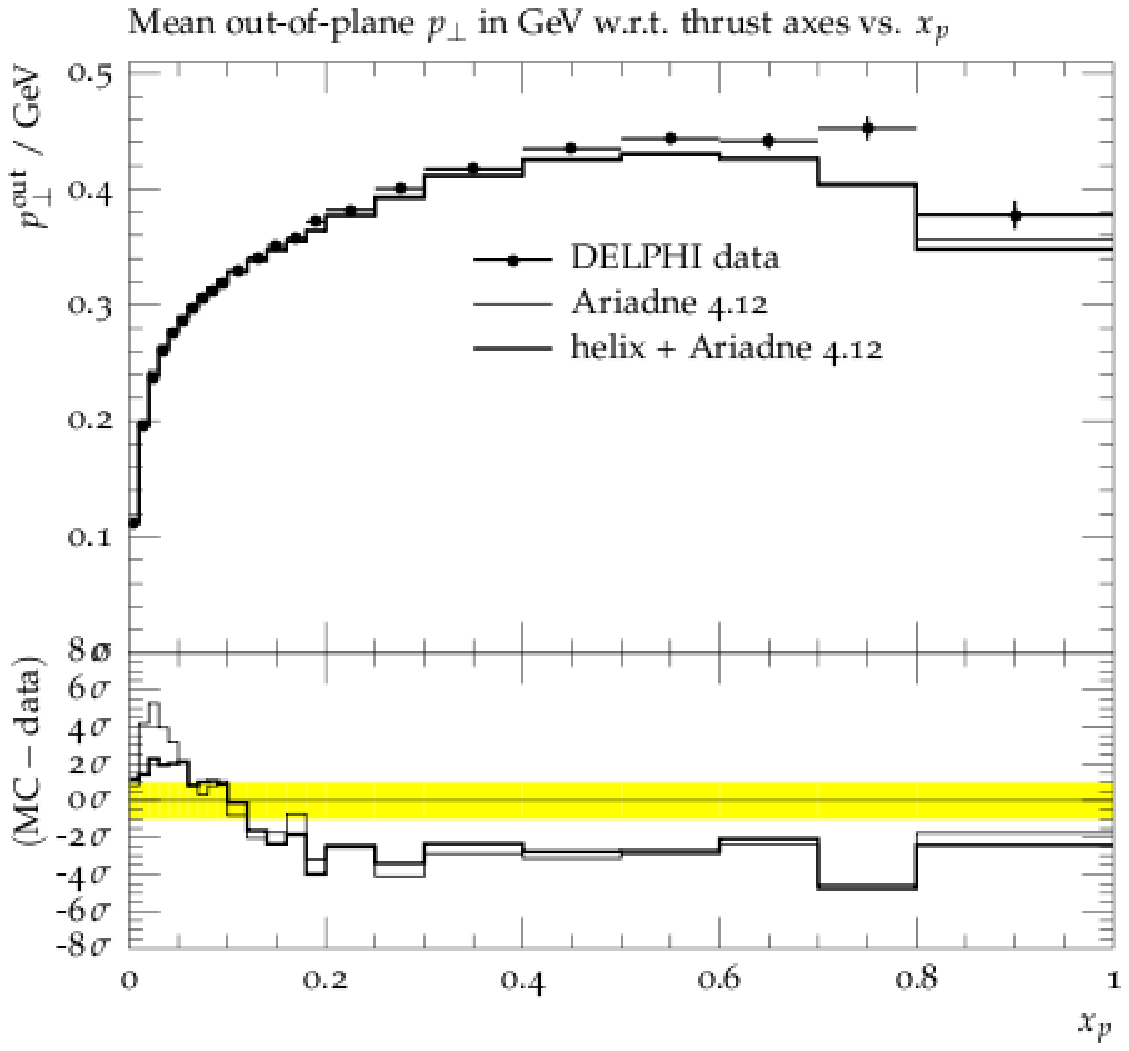}
\includegraphics[width=0.5\textwidth]{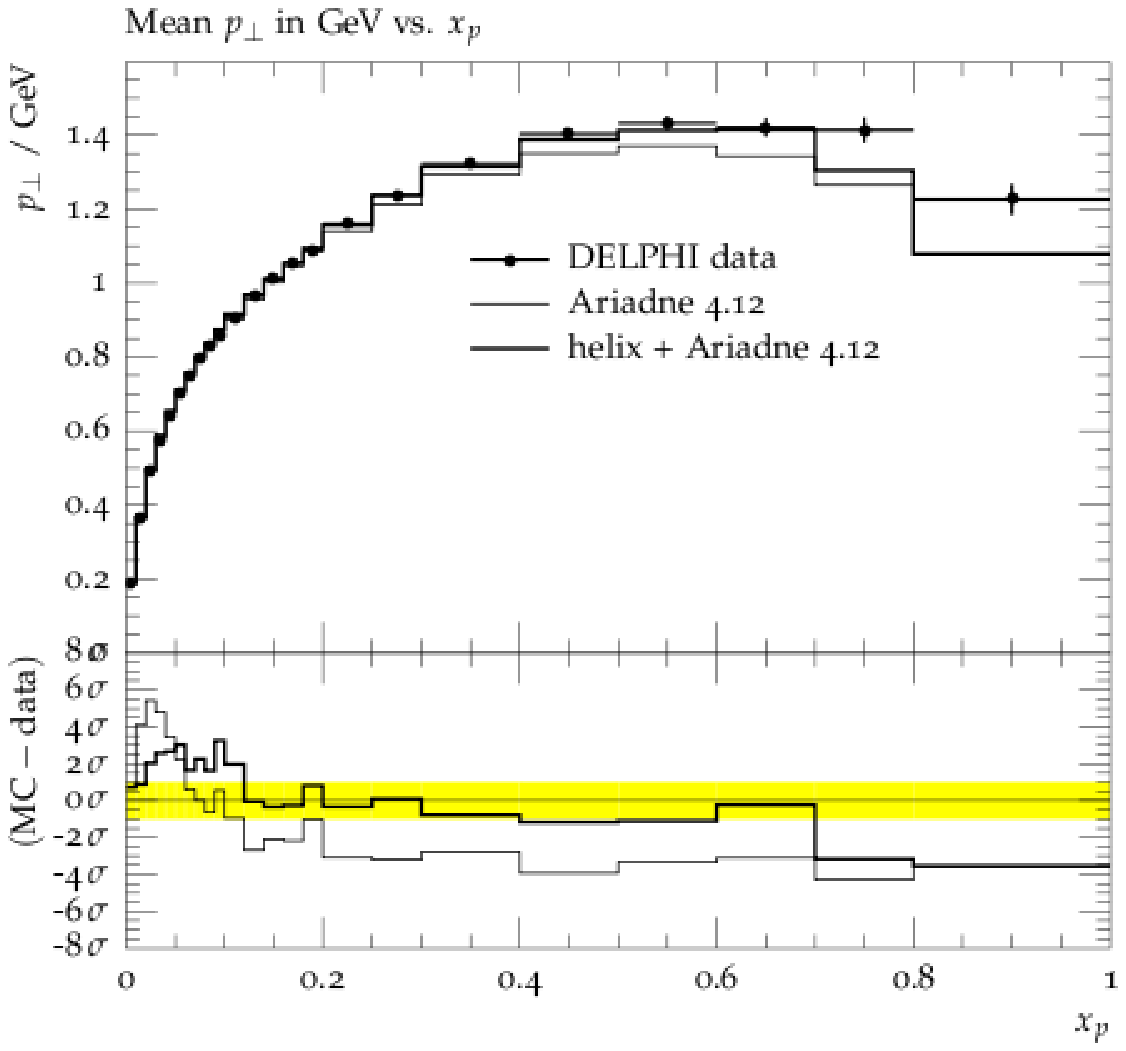}
\includegraphics[width=0.5\textwidth]{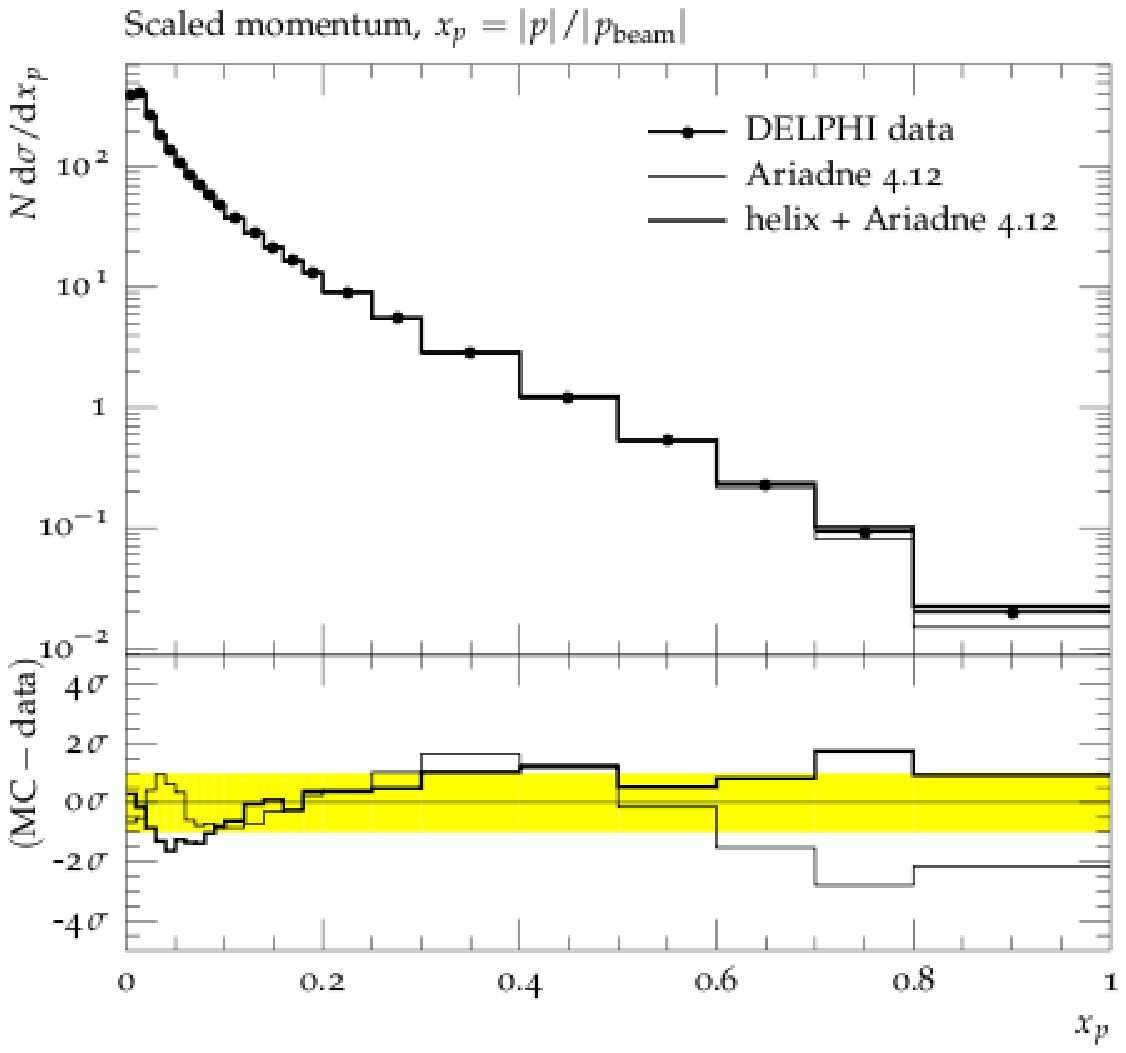}
\includegraphics[width=0.5\textwidth]{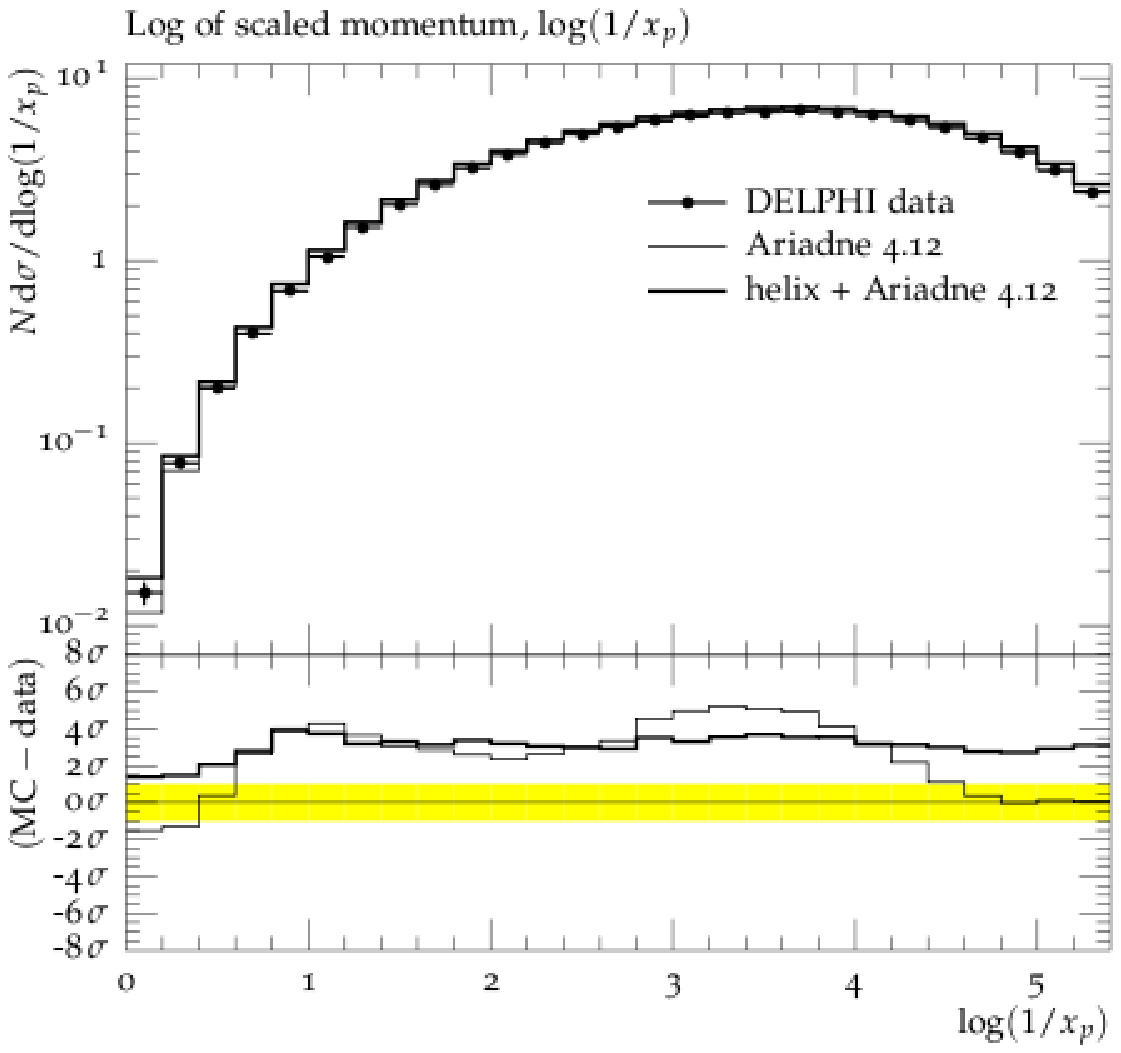}
\caption{Comparison of inclusive charged particle distribution measured by DELPHI
 and MC simulation based on Ariadne parton shower, using standard
 fragmentation ('Ariadne 4.12') or helix string model ('helix + Ariadne 4.12').}
\label{fig:incl_ari} 
\end{figure}

 With implementation of the helix string model, the string fragmentation effectively
 looses one degree of freedom, and the additional constraint translates into modified
 transverse momentum distribution. The change is visible in the low $p_T$ part of the transverse
 momentum distribution ( $<$ 1 GeV/c ), and it is supported by the data
 (see plots on top of Figs.~\ref{fig:incl_pt},~\ref{fig:incl_ari}).

 Even more significant for the model validation is the improvement in
 the dependence of the average transverse momentum on the scaled momentum of the hadron
 (middle plots in Figs.~\ref{fig:incl_pt},~\ref{fig:incl_ari}). This may well be the
 best evidence in favour of the existence of inherent correlations between transverse
 and longitudinal components of hadron momenta.

   The positive effect of the helix string modelling is seen also on a number of
 event shape variables (most notably on Major, minor; hemisphere broadening; differential
 jet rates). The overall $\chi^2$ for the whole set of distributions used in the tune
 is given in Table~\ref{table:chi2}, and the numbers clearly indicate the preference
 of the data for the helix string model, when compared with the standard fragmentation
 of the Lund string. For completeness, the $\chi^2$ is quoted also for the set of
 identified particle rates, and the b-fragmentation function, even though these
 distributions are not included in the present tune. The tuned parameter values are
 quoted in the Appendix.

\begin{table}[h]

\begin{tabular}{| l || c | c || c | c |} 
\hline
  Data set  &  Pythia \cite{holger} & helix + Pythia & Ariadne & helix + Ariadne \\  
\hline
 inclusive spectra   &  & & &  \\
 + event shapes & 4075 & 2453 & 2453 & 1489 \\
 $N_{bin}=619$  &  & & &  \\
\hline
 ident.part.rates & & & & \\
 + b-fragmentation & 444 & 669(*) & 614(*) & 586(*) \\
 $N_{bin}=47$  &  & & &  \\ 
\hline
\end{tabular}
\caption{ Sum (over all bins) of $\chi^2$ difference between data and models. The 'Pythia/Ariadne' labels distinguish
 between Pythia 6.421 pT-ordered parton shower, and Ariadne 4.12 parton shower. (*) distributions
 not included in the tune.}
\label{table:chi2} 

\end{table}

\begin{figure}[bth]
\includegraphics[width=0.5\textwidth]{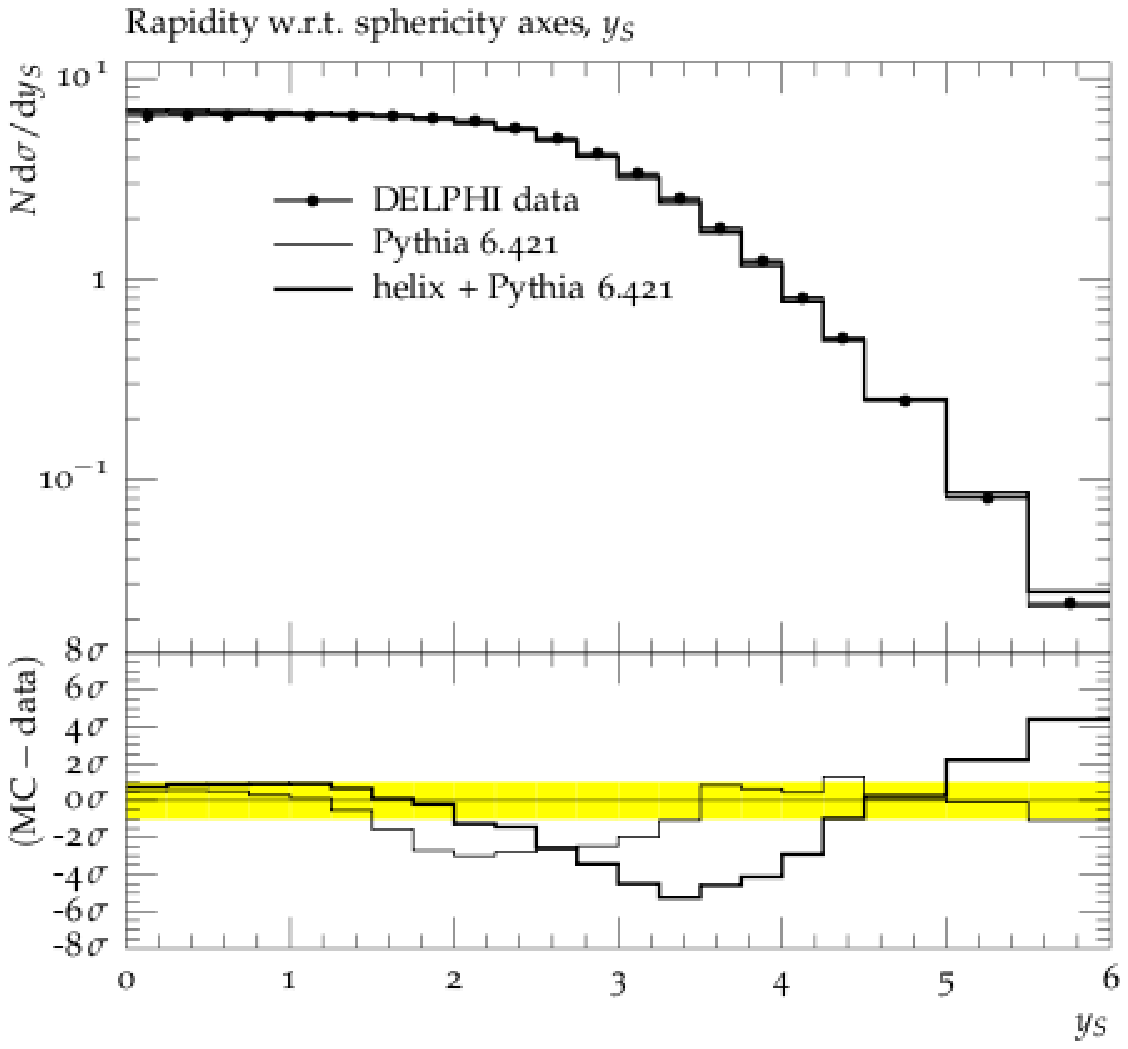}
\includegraphics[width=0.5\textwidth]{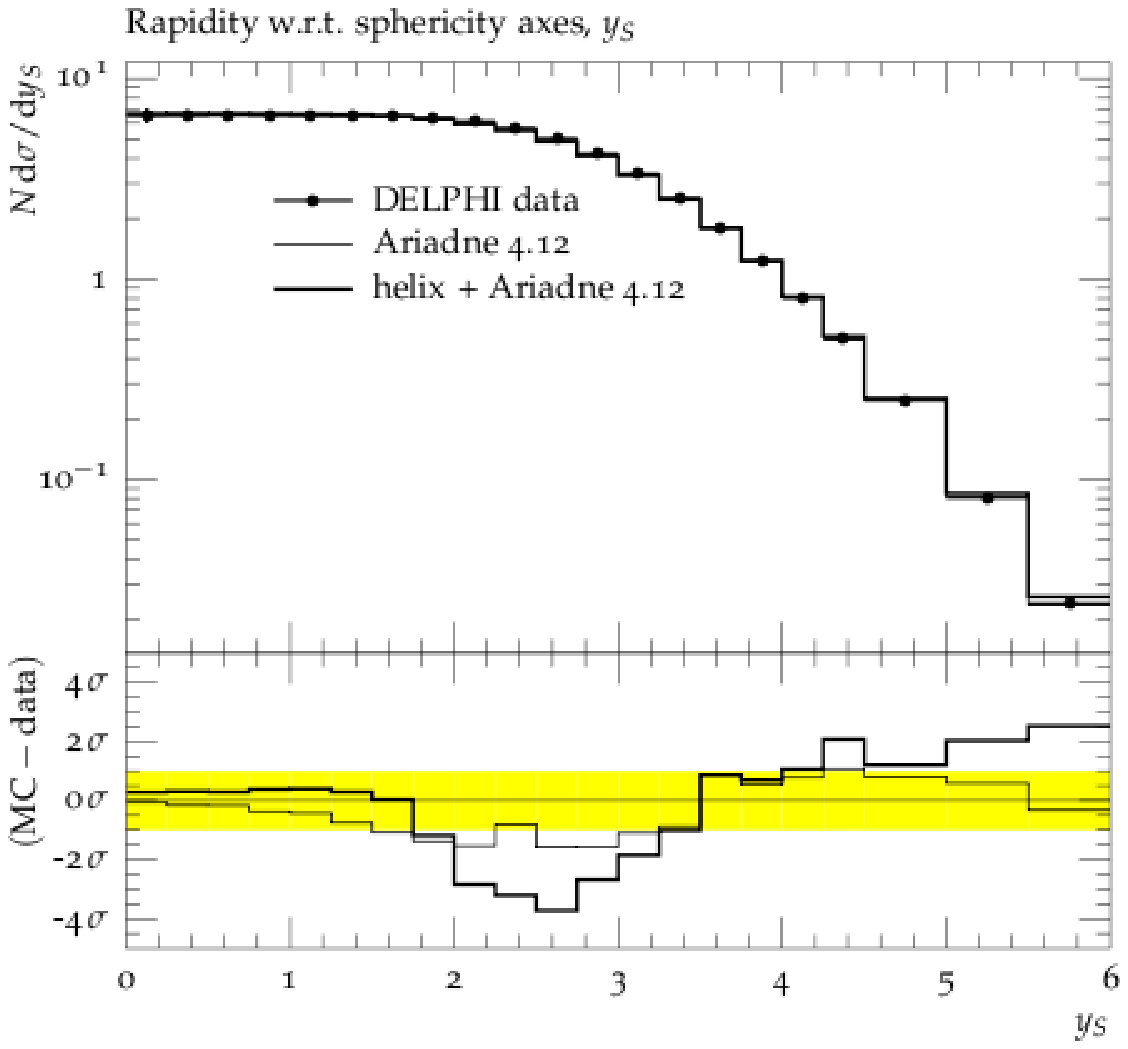}
\caption{Comparison of rapidity distribution ( w.r.t. sphericity axis )
 as measured by DELPHI, and MC simulation based on Pythia pT-ordered shower (left) and 
 Ariadne parton shower (right), for standard fragmentation (dotted line)
 and for helix string model (solid line).}
\label{fig:ys} 
\end{figure}

  Despite the overall improvement of the data description due to the helix string model, 
 there are few distributions where a degradation is observed.  There is some negative impact
 on the rapidity distributions (Fig.\ref{fig:ys}), and 
 some degradation of the description of the scaled momentum is observed in the modelling
 based on Pythia pT-ordered parton shower (bottom plots
 in Fig.\ref{fig:incl_pt}). The later is not confirmed in the study 
 running Ariadne parton shower (Fig.\ref{fig:incl_ari}),
 and there are some indications the effect may be related to the (untuned) charm quark fragmentation function.

\clearpage

\section{Conclusions}
      
    A significant improvement in the description of hadronic $Z^0$ data is achieved
 with the help of modified string model which assumes the gluon field is ordered in a helix-like
 structure with pitch proportional to the energy density stored in the field. 

   The improvement is achieved despite the minimal tuning setup which does not
 readjust the parameters of the heavy quark fragmentation. This means there is definitely
 room for further improvement of the tune. 

    Tuned helix string parameters suggest a helix radius r $\sim$ 0.4 GeV/c (rather well constrained)
 and a helix winding of S $\sim$ 0.7 rad/GeV ( with large uncertainty ).

\section*{Acknowledgements}

   The author is grateful to the Rivet and Professor Collaborations for providing 
 efficient tools for model development and validation. 
          
\clearpage

%
%
 

\begin{footnotesize}

\end{footnotesize}


\section*{Appendix}

  \subsection*{i/ Tuning of the model with Pythia p$_T$-ordered parton shower }

    The setup of the helix string model tune corresponds to the Professor flavour and p$_T$ tune \cite{prof_tune} with the
 exception of PARJ(47), fixed at 0.873 instead of 0.8. The change is done in order to facilitate
 comparison of the helix string tune with the updated Professor fragmentation tune (\cite{holger}).  
  
    Tuned parameter values and fit quality are given in Table~\ref{table:pt}.    

 \begin{table}[h]
\begin{center}
\begin{tabular}{|c|c|} 
\hline
Parameter &  Tuned value \\
\hline 
PARJ(41)  &  0.084 $\pm$ 0.12     \\
PARJ(42)  & 0.375 $\pm$ 0.19  \\
PARJ(81)  & 0.237 $\pm$ 0.025   \\
PARJ(82)  & 0.65 $\pm$ 0.59   \\
PARJ(102) & 0.362 $\pm$ 0.032      \\
PARJ(104) & 0.510 $\pm$  0.23   \\
\hline
 $\chi^2/(n_{bin}-n_{d.o.f}) $  &   4. \\
\hline
\end{tabular}
\caption{ Results of the 6 parameter tune of the helix string model combined with Pythia 6.421 pT-ordered shower. }
\label{table:pt}
\end{center}
\end{table}

  \subsection*{ii/ Tuning of the model with Ariadne parton shower }

    The setup is similar to the one used in the DELPHI tune \cite{tuning} with following modifications: the detailed identified particle
 rate steering is not available, Bose-Einstein effect simulation is switched off ( MSTJ(51)=0 ), and radiative corrections are
 included ( MSTJ(107)=4 ). The tuning is performed twice, for the standard Pythia fragmentation, and for the helix string
 fragmentation. The tuned parameter values and fit quality are given in Tables~\ref{table:ari0},~\ref{table:arihel}.

\begin{table}[h]
\begin{center}
\begin{tabular}{|c|c|} 
\hline
Parameter &  Tuned value \\
\hline 
PARA(1)  & 0.240 $\pm$ 0.021   \\
PARA(3)  & 0.537  $\pm$ 0.22  \\
PARJ(41)  & 0.41 $\pm$ 0.13     \\
PARJ(42)  & 0.85 F  \\
PARJ(21) & 0.362 $\pm$ 0.032     \\
\hline
 $\chi^2/(n_{bin}-n_{d.o.f}) $  &   4. = 2453/(619-5) \\
\hline
\end{tabular}
\caption{ Results of the 5 parameter tune of Ariadne 4.12 with standard Pythia fragmentation. }
\label{table:ari0}
\end{center}
\end{table}

\begin{table}[h]
\begin{center}
\begin{tabular}{|c|c|} 
\hline
Parameter &  Tuned value \\
\hline 
PARA(1)  & 0.264 $\pm$ 0.024   \\
PARA(3)  & 1.39  $\pm$ 0.35  \\
PARJ(41)  & 0.540 $\pm$ 0.090     \\
PARJ(42)  & 0.576 $\pm$ 0.108  \\
PARJ(102) & 0.429 $\pm$ 0.028     \\
PARJ(104) & 0.682 $\pm$ 0.165     \\
\hline
 $\chi^2/(n_{bin}-n_{d.o.f}) $  &   2.4 = 1489/(619-6)\\
\hline
\end{tabular}
\caption{ Results of the 6 parameter tune of Ariadne 4.12 with helix string fragmentation. }
\label{table:arihel}
\end{center}
\end{table}

\end{document}